\documentclass[aps,pre,onecolumn,superscriptaddress,floatfix,tightenlines,showpacs,notitlepage]{revtex4-1}
\usepackage{graphicx}
\usepackage{bm}
\usepackage[normalem]{ulem}
\usepackage{amsfonts}
\usepackage{color}
\usepackage[dvipsnames]{xcolor}
\usepackage{amsmath}    
\usepackage{epsfig}
\usepackage{subfigure}  
\usepackage[breaklinks,colorlinks = true,linkcolor = blue,urlcolor=blue,citecolor=blue]{hyperref}
\usepackage{amssymb}
\usepackage{comment}

\input{symdef.inc}

\begin{document}
	\title{Elasticity of fibres prefers the chaos of turbulence}
	\author{Rahul K. Singh}
	\email{rksphys@gmail.com}
	\affiliation{\oist}
	
	\begin{abstract}
		The dynamics of fibres, modelled as a sequence of inertial beads linked via elastic springs, in turbulent flows is dictated by a non-trivial interplay of their inertia and elasticity. Such elastic, inertial fibres preferentially sample a three-dimensional turbulent flow in a manner qualitatively similar to that in two-dimensions [Singh \textit{et al.}, Phys. Rev. E \textbf{101}, 053105 (2020)]. Inertia and elasticity have competing effects on fibre dynamics: Inertia drives fibres away from vortices while elasticity tends to trap them inside. However, both these effects are reversed at large values. A  large inertia makes the fibres sample the flow more uniformly while a very large elasticity facilitates the sampling of straining regions. This complex sampling behaviour is further corroborated by quantifying the chaotic nature of sampled flow regions. This is achieved by evaluating the maximal Lagrangian Lyapunov Exponents associated with the flow along fibre trajectories. 
	\end{abstract}
	
	\maketitle

Turbulent flows are ubiquitous in nature and are responsible for numerous transport phenomena that help sustain life on earth. The transport of heat by atmosphere and oceans~\citep{Ganachaud2000,Ferrari2011,EricOceans2014,Su2018,Han2017} and the formation of raindrops and clouds governs climate dynamics at the planetary scales~\citep{Falkovich2002,Lau2003,Gabrowski2013,Lanotte01}, while that of planktons, sediments and nutrients by oceans and rivers is vital for agriculture and marine life~\citep{Durham2013,Stocker2015,Vercruysse2017}. Turbulent transport also lies at the heart of dispersion of ash and pollutants from volcanic eruptions and industrial exhausts which can be laden with particles of varying sizes. Particulate matter transported by turbulence is often small in size, which makes their study relatively simple~\cite{Falkovich2001,Bec2006,SSR2014a,SSR2014b,Voth2017,Brandt2022}. Objects transported by turbulent flows can, in general, also be interacting leading to
different forms of aggregation~\cite{Pumir2016,Bec2016,SSRSticky13,Lanotte2023}; they can also have more complex dynamics
owing to deformability~\cite{Soldati2017,Verhille2022} or intrinsic anisotropy such as in ice crystals of
cold clouds~\cite{Klett1997,Wang2013,Prateek2020,Gustavsson2022,Prateek2024}; and can have characteristic length scales
longer than the small-scales of turbulence which not only makes them
anisotropic and deformable, but also extensile. Common examples include ice crystals in clouds~\cite{Klett1997,Wang2013}, pollen dispersion by air~\citep{Sabban2011}, soot emissions~\cite{Kohler2011}, fibres in industries such as papermaking~\cite{Lundell2011,Shankar2020}, textiles manufacturing~\cite{Wegener06}, and fabric design and manufacturing~\cite{Jarecki06,Stocker12,Wilczek18}. 

Various aspects of the dynamics of anisotropic, elongated deformable fibres in turbulent flows such as their stretching, deformation, tumbling~\cite{Verhille14,Verhille16,Verhille18,Gay18,Settling21}, clustering~\cite{Marchioli19}, and buckling~\cite{Bec18} are now very well understood. Rigid, inflexible fibres have been shown to be reliable proxies for obtaining two-point statistics of turbulence~\cite{Rosti18,Rosti19}. Elastic, extensible and fully flexible fibres are stretched out by both vortical and straining regions in 3D turbulence~\citep{RoyalChains} and can break into smaller fragments~\cite{Picardo2021}. Inertial fibres in turbulent flows settle under gravity with an enhanced speed whose fluctuations depend on inertia as its inverse-square but remain independent of the fibre elasticity~\citep{Settling21}. These fully-flexible, elasto-inertial fibres preferentially sample the straining regions in two-dimensional (2D) turbulence~\citep{SinghPRE} in contrast to inertia-less fibres that prefer the vortical regions~\citep{PicardoPRL}. 
In this work, we focus on the Lagrangian statistics and dynamics of such long, extensile fibres and show how the the complex interplay of their inertia and elasticity influences the way elasto-inertial fibres preferentially sample a 3D turbulent flow. We also relate this picture of preferential sampling to the chaotic nature of the flow regions visited by the fibres.

We address the above problems by generating a turbulent carrier flow via the direct numerical simulations of the Navier-Stokes Eqns.~\ref{NS} complemented by a simultaneous evolution of fibre trajectories governed by Eqns.~\ref{Fibre},~\ref{Fibcm}. 
\begin{align}
	 	\frac{\partial \bfu}{\partial t} + \bfu \cdot \boldsymbol{\nabla} \bfu = - \frac{1}{\rhof} \boldsymbol{\nabla} p  + \nu \boldsymbol{\nabla}^2  \bfu + \textbf{f}			\label{NS}
\end{align}
 	where $\bfu$ is the incompressible fluid velocity, i.e. $\boldsymbol{\nabla} \cdot \bfu = 0$ so that the density $\rhof = 1$ can be set to a constant value. A statistically stationary, turbulent state is maintained by energy injection at large scales which is enforced by the forcing {\bf f}. This energy is dissipated away at small scales by the (kinematic) viscosity $\nu$. More details on simulations are provided later.
 	 The flow, in a statistically stationary turbulent state, is then seeded with $(10^4)$ fibres whose modelling and equations of motion are discussed in the following section.

\section*{The Elasto-inertial Chain: a minimal model for fibres}

We model our fibres as $\Nb$ identical, inertial beads (spheres) that interact with their nearest neighbors via (phantom) elastic links. Our inertial beads are sub-Kolmogorov scale, i.e. $a \ll \eta$, characterised by a Stokesian relaxation time $\taup= \frac{2\rho a^2}{9\rhof\nu}$, where $\rho$ is the density of the bead while the elastic links are modelled as FENE (finitely extensible nonlinear elastic) springs with the constant for the $j$-th link is given as $f_j=(1-|\bm{r}^2_j|/r^2_{\rm m})^{-1}$ such that $\bm{r}_j = {\bf x}_{j+1} - {\bf x}_{j}$ is the instantaneous separation vector between the $j$-th and the $(j+1)$-th bead and $r_{\rm m}$ is the maximum extension of any link. 
These elastic springs are characterised by a relaxation time $\taue$ (which gives an effective elastic time scale $\taue \Nb (\Nb + 1)/6$ for the fibre~\cite{Collins2007}). One can thus write an equation of motion for a fibre in terms of the individual spring vectors and its $\com$  as:

\begin{align} 
\taup\ddot{\bm{r}}_j &= \left[\bu{j+1}-\bu{j} - \drj \right] + A\left [{\bm \xi}_{j+1}(t)-{\bm \xi}_j(t)\right ] +\frac{1}{4\taue}\left (f_{j-1}\bm{r}_{j-1}-2f_{j}\bm{r}_{j}+f_{j+1}\bm{r}_{j+1}\right )  \label{Fibre}\\
	\taup \xcdd &=
	\left(\frac{1}{\Nb}\sum_{j=1}^{\Nb}{\bu{j}}-\xcd \right)+\frac{A}{\Nb}\sum_{j=1}^{\Nb}{\bm
		\xi}_j(t).  \label{Fibcm} 
\end{align}

where ${\bm \xi}_j(t)$ are independent white noises which are introduced to set the equilibrium length $r_0$ of the elastic links (in the absence of flow). The noise amplitude is chosen accordingly as $A^2 = r^2_0/6\taue$. The entire fibre then has equilibrium and maximum end-to-end extensions of $R_0 = r_0 \sqrt{\Nb-1}$ and $R_{\rm m} = r_{\rm m} (\Nb-1)$, respectively. We note that the precise method of enforcing the equilibrium length is not crucial to our study. Setting the equilibrium length in a different manner such as a spring force of the form $f_j (r_j - r_0)$ does not affect the qualitative dynamics of the fibres in flow~\cite{RoyalChains}. Finally, the time-scales $\taup$ and $\taue$ allow us to define non-dimensional numbers in terms of analogous (small-scale) quantities of the turbulent carrier flow, namely the Kolmogorov time scale $\tau_\eta \equiv \sqrt{\nu/\epsilon}$ where $\epsilon$ is the mean energy dissipation rate of the flow. The dynamics of our fibres are thus completely determined by the Stokes number St $\equiv \taup/\tau_\eta$ (a measure of the inertia), and the Weissenberg number Wi $\equiv \frac{\Nb(\Nb+1)\taue}{6\tau_\eta}$ (a measure of elasticity). Here, we have used the mapping proposed by \cite{Collins2007} to estimate the effective relaxation time of the entire fibre from $\taue$ (which corresponds to individual links).

Note that we consider only the extremely dilute limit of a single fibre so that the motion of the fibre does not affect the global flow. A large number of fibres are simulated only to obtain good statistics. Thus, we neglect any back reaction of the fibres on the flow. Moreover, different segments of the same fibre can interact via hydrodynamic interactions of the disturbed flow. In fact, this can be accounted for by including inter-bead hydrodynamic interactions (HI) as has been done in polymer models~\cite{Jendrejack2002,Schroeder2004}. However, the effect of HI on fibre dynamics is minimal when the fibre is stretched out to very long lengths by the flow. This is true for our fibres whose lengths lie in the deep inertial range. We therefore ignore HI. We also ignore excluded volume interactions on similar grounds. Thus, ours is a minimal model of a fibre, adopted in order to reveal the fundamental interplay between elasticity, inertia, and turbulent advection.

It is worth mentioning that other models of fibre dynamics, particularly in the context of turbulent transport, have also been studied in detail~\cite{Brouzet_polymer,Verhille_3dconf_fiber,Brandt_fiber,Bec_Chain}. However, earlier works on low Reynolds number flow show~\cite{Marchetti} that the bead-spring approach to fibres is an important framework~\cite{FilamentPRL,Schlagberger,Llopis,Delmotte} because of the limitations of models based on slender-body theory~\cite{cox_1970,Xu}.

\begin{figure*}[!ht]
	\includegraphics[width=0.8\textwidth]{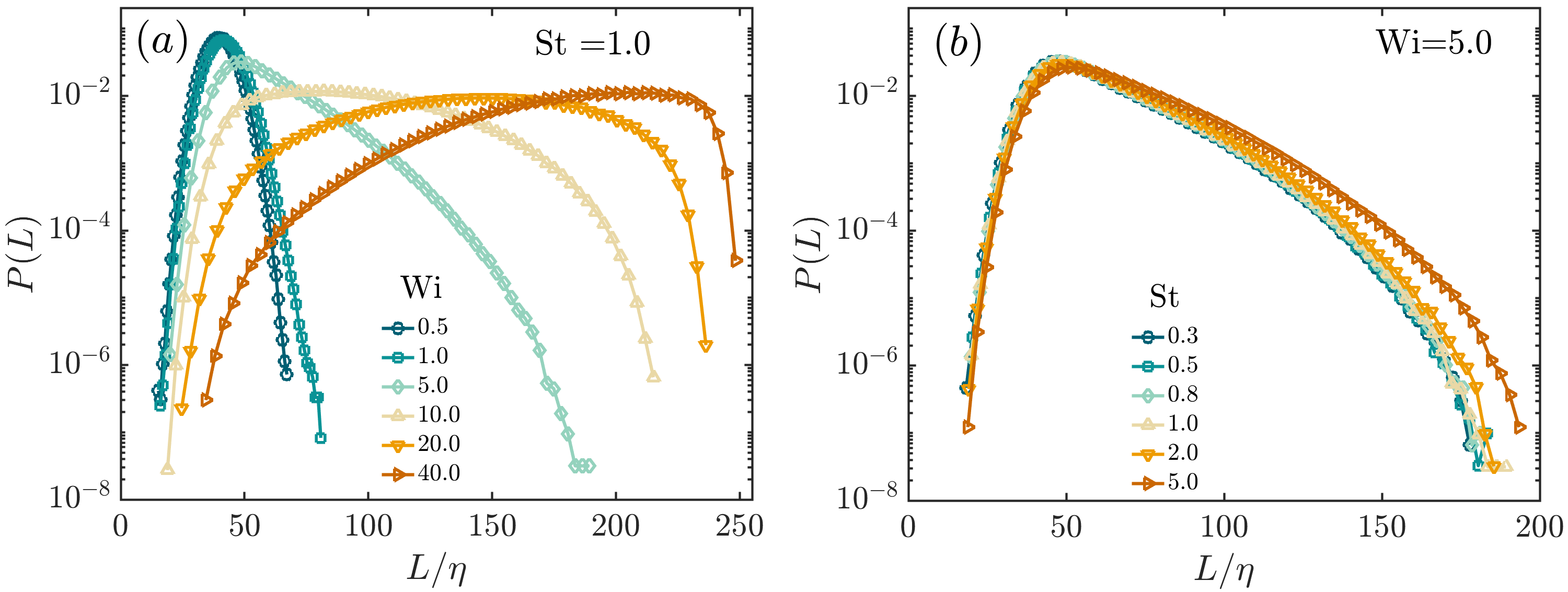}
	\caption[\textbf{Filament Lengths in 3D}]{Probability Distribution Functions (PDFs) of total lengths $L/\eta$) of our elasto-inertial fibres as a function of (a) Wi when St = 1 and (b) St when Wi = 5.0. A large Wi indicates a more extensible fibre implying that typical lengths will be longer. This is seen in the the PDFs of $L$ peaking at larger values for large Wi in panel (a). Fibre inertia has a minimal effect on its stretching and the PDFs collapse for different St in panel (b). At large inertia, fibres cannot follow the flow and resist the tendency of springs to relax back more. This results in the broadening of the right tails of the pdfs at large St.}
	\label{LPDF}
\end{figure*}

\section*{Simulations}
We solve the three-dimensional, incompressible Navier-Stokes Eqns.~\ref{NS} to drive the flow to a
statistically stationary state by using a constant energy injection scheme. A choice of $\nu = 10^{-3}$ allows us to obtain a Taylor-scale Reynolds number $Re_\lambda \approx 200$. We use a de-aliased pseudo-spectral algorithm that spatially discretizes the Navier-Stokes equations on a 2$\pi$ periodic cubic box into $N^3 = 512^3$ collocation points while a second-order slaved Adams-Bashforth scheme is employed for evolving in time~\cite{James17}. We simultaneously solve the equations of motion of the fibres~\ref{Fibre},~\ref{Fibcm} using a second-order Runge-Kutta scheme for the deterministic terms while the noise term is evolved using the Euler-Maruyama method~\citep{Ottinger1996,Settling21}. The fluid velocity $\bu{}$ obtained on the regular periodic grid is interpolated, using trilinear interpolation to obtain the fluid velocity $\bu{j}$ at the bead positions. We evolve an ensemble of $10^4$ identical fibres each with an equilibrium end-to-end extension of $R_0 = 15.2\eta$, and a maximum length $L = 270\eta$. Each fibre comprises of $\Nb = 10$ beads interacting with the nearest neighbours via $\Nb -1 = 9$ links. We span almost a decade in both fibre inertia $0.1 \leq St \leq 8.0$, and elasticity $0.5 \leq Wi \leq 40$.

\section*{Filament Lengths}
\label{Total Lengths}

To understand fibre dynamics, we begin by discussing first how the length and the shape of the fibres are determined by their inherent inertia and elasticity. A fibre with its equilibrium length $r_0$ in the inertial range of scales, i.e. $L_0 \gg r_0 \gg \eta$, where $L_0$ is the scale of forcing, feels the full roughness of the turbulent carrier flow. Different beads of a fibre experience completely different drag forces resulting from very different local fluid velocities. This causes the fibres to stretch and acquire dynamically varying shapes. A less elastic (small Wi) fibre has a larger effective spring constant and is more stiff compared to a large Wi fibre that can readily stretch to very long lengths. Expectedly, the length of a small Wi fibre does not deviate appreciably from its equilibrium value. This becomes clear from the statistics of the instantaneous total contour length of a fibre $L = \sum_{i=1}^{\Nb-1} r_i$, where $r_i$ is the length of the individual springs (or the separation between successive beads). We show in Fig.~\ref{LPDF}(a) the probability distribution functions (pdfs) of $L/\eta$ for varying degrees of fibre elasticity ata fixed inertia of St = 1.0. The distributions are clearly sharply peaked with a small spread when Wi is small. As Wi is increased, fibres begin to stretch more, resulting in a non-zero probability for large $L$ and making the distributions more uniform. Highly elastic fibres are stretched out maximally by the flow so that the pdfs for very large Wi develop a peak close to (but slightly lesser than) the maximum possible length $L_{\rm m} = 270\eta$. The stretching of fibres, however, remains largely independent of their inertia as seen from Fig.~\ref{LPDF}(b). The $L/\eta$-pdfs collapse for different St except for very large inertia. Fibres with large inertia can resist the relaxation of elastic springs to their equilibrium lengths more. Additionally, they are unable to follow the local flow closely and thus once stretched, they remain so for longer times. This means that $L$-pdfs have marginally wider right tails when St is large as seen in Fig.~\ref{LPDF}(b).

\begin{figure}[!ht]
	\centering
	\includegraphics[width=0.5\columnwidth]{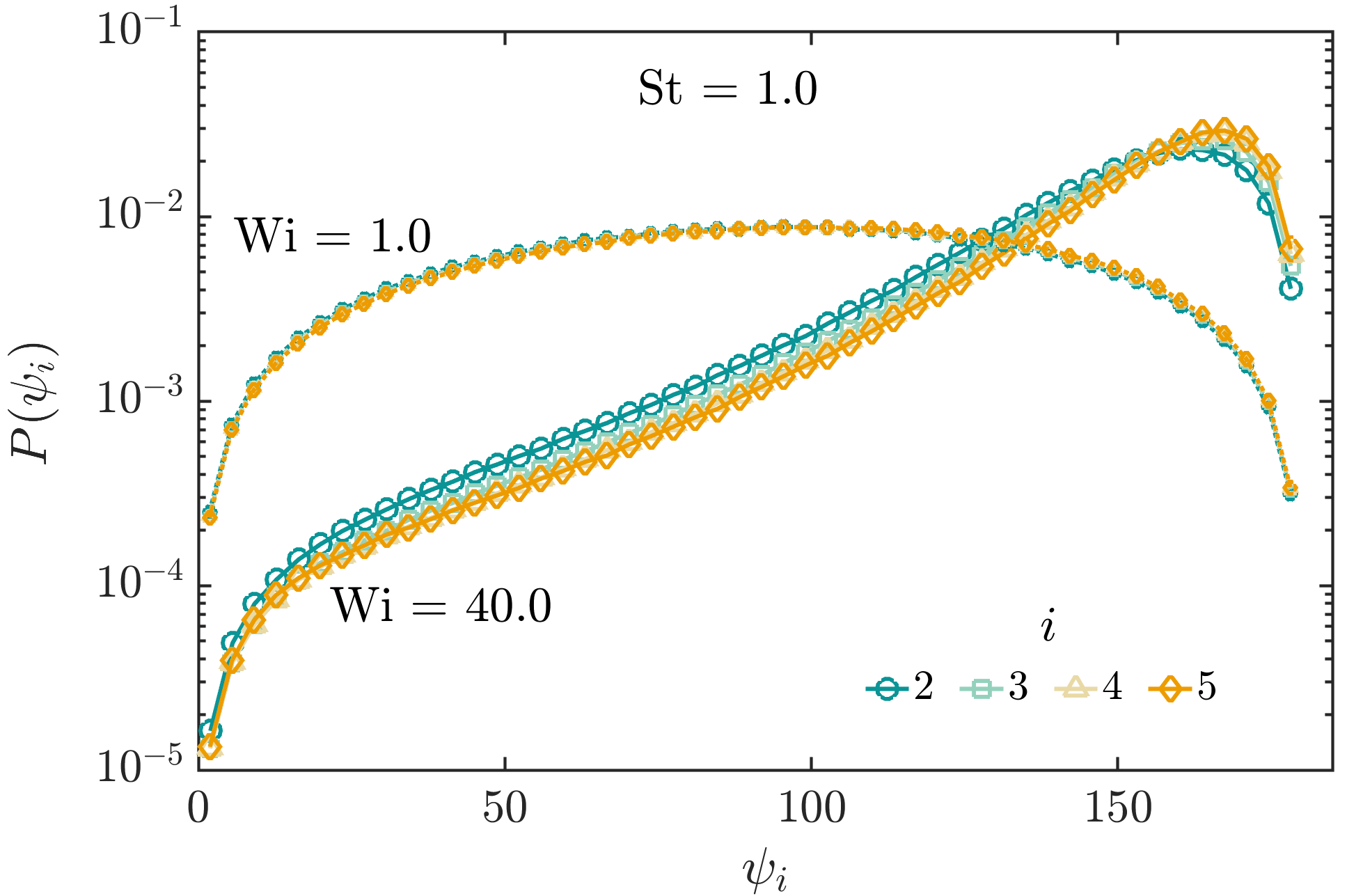}
	\caption[\textbf{Distribution of Hinge Angles}]{PDFs of the first four hinge angles for two different instances of elasticity Wi $= 40$ (solid lines, big markers) and, Wi $= 1.0$ (dotted lines, small markers). The statistics of hinge angles are unique only for the beads 2-5 due to the head-tail symmetry of the fibre model. A more elastic fibre with Wi = 40 is typically more stretched out making the hinge angles typically large. A less elastic fibre with Wi = 1 shows no such preference and has a more uniform distribution of hinge angles.}
	\label{HAF}
\end{figure}

\section*{Shape of the Filaments: Hinge Angles}

Our minimal model of fibres doesn't incorporate any cost of bending (which was shown to have a minimal effect on the various fibre dynamics in~\citep{RoyalChains,Settling21}) rendering them fully flexible. In a fully flexible fibre, the angles $\psi_i, i = 2,3,...,\Nb-1$ between successive links (springs) can take values in $[0, 180^\circ]$, where $\psi_i$ is defined as:
\begin{align}
	\cos \psi_i = \frac{\rj{i-1} \cdot \rj{i}}{ \modls{\rj{i-1}}  \modls{ \rj{i}}}.
\end{align}
The first and the last beads do not form any hinge angles as they are connected to only one spring each.
A fibre with $\psi_i = 180^{\circ}$ $\forall$ $i$ has an end-to-end length equal to its total length:
\begin{align}
\modls{\sum_{i=1}^{\Nb}\rj{i}} = \sum_{i=1}^{\Nb}  \modls{\rj{i}},		\qquad \psi_i = 180^{\circ} \quad \forall i
\end{align}  
Similarly, fibres with smaller hinge angles will have their end-to-end lengths less than their total length. Hinge angles can, therefore, be used a proxy measure of the shape of the fibres. Now, we see that our model Eqn.~\ref{Fibre} imply that the fibres have a head-tail symmetry, i.e. the equation retains its form under the transformation $\rj{j} \rightarrow -\rj{j}$. This implies that the dynamics of $i$-th bead ($i$-th spring) is statistically identical to that of $(\Nb - i + 1)$-th bead ($(\Nb - i )$-th spring). As a result, the statistics of $\psi_i$ are unique only for first $\Nb/2$ $((\Nb+1)/2)$ beads for $\Nb$ even (odd). Therefore, we only show the pdfs of the first four (unique) hinge angles in Fig.~\ref{HAF}  for Wi = 40 (solid lines, large markers) and Wi = 1 (dotted lines, small markers). Both cases have the same inertia of St = 1.0. Clearly, fibres with large elasticity are more likely to exhibit large hinge angles as they are stretched out by the carrier flow. This stretching makes large hinge angles more probable so that the pdfs peak close to $180^\circ$ when Wi is large. However, at a small Wi, the stiffness of the springs does not allow the fibres to stretch. As a result, such fibres experience more correlated drag forces along their contours and are, thus, less likely to be completely stretched out. (Fibres are stretched more when drag forces on the beads are very different.) These small fibres thus show no particular preference of hinge angles and exhibit a rather uniform distribution of these angles. This is seen in the almost flat pdfs of $\psi_i$ at small Wi in Fig.~\ref{HAF}.

\section*{Preferential Sampling}
We have now clearly established how the shape and size of fibres depends strongly on their inertia and elasticity. This now helps us to address the question of preferential sampling: How does our model fibre sample a 3D turbulent flow and how does it depend on the interplay of inertia and elasticity? It is already known that an inertia-less fibre (one comprising non-inertial, tracer beads) preferentially samples the vortical regions of a turbulent flow both in 2D and 3D, but owing to different underlying reasons~\cite{PicardoPRL,RoyalChains}.
However, inertial fibres tend to avoid the vortical regions and show a preference for straining zones in 2D turbulence~\citep{SinghPRE}. We now discuss how our elasto-inertial fibres sample a 3D turbulent flow. 

One way to quantify preferential sampling is via the geometry of the local flow. Consider the $\q$-criterion (analogous to the Okubo-Weiss parameter in 2D) which describes whether the local flow is vortical or extensional. The local, instantaneous $\q$-value can be obtained using the velocity gradient $\bm{A}({\bf x},t)= \boldsymbol{\nabla} \bfu$ at a location ${\bf x}$ and time $t$ as:
\begin{align}
	\q = \tau^2_{\eta} \frac{\omega^2 - S:S}{2} \ ; \  \omega = \frac{A - A^\intercal}{2},  S = \frac{A + A^\intercal}{2}
\end{align}  
where $\omega$ is a measure of local rate of rotation (vorticity) and $S$ is a measure of the local strain. Clearly, the sign of $\q$ indicates a vorticity ($\q > 0$) or a strain ($\q < 0$) dominated region of the flow. 

\begin{figure}[!ht]
	\centering
	\includegraphics[width=.5\columnwidth]{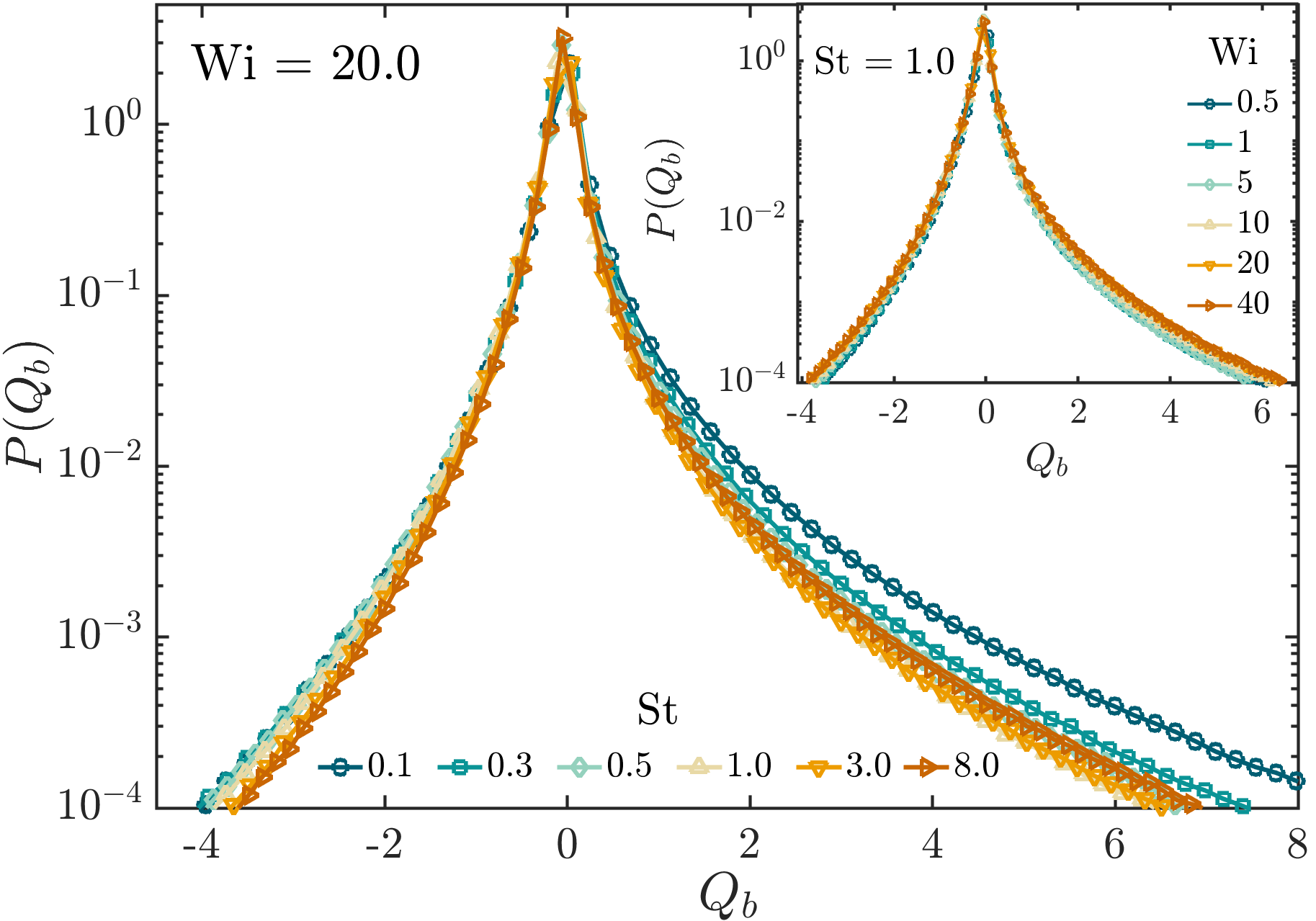}
	\caption[\textbf{Preferential Sampling in 3D by Inertial Filaments}]{PDFs of the $\q$ values measured at the bead positions ($\qb$) of a fibre for a fixed Wi and varying St (main), and for fixed St and varying Wi (inset). (Main) Wider right tails indicate preferential vortical trapping of fibres. As inertia increases, fibres are expelled from the vortices resulting in shrinking of the right tails. (Inset) The dependence of preferential sampling on fibre elasticity. Long, elastic fibres tend to be trapped for longer in both vortical and straining regions. Hence, the right as well as left tails of the PDFs are widest when elasticity is largest (for fixed inertia).}
	\label{Qpdf1}
\end{figure}
With this definition in place, we track the Lagrangian history of the $\q$-values at all the bead positions ($\q_b$) for all fibres.  The distribution of these  $\q$-values as a function of inertia and elasticity is shown in Fig.~\ref{Qpdf1}. In the limit of small inertia, i.e. when St $\to 0$, one expects a close to tracer behaviour when the $\q$-pdfs show a maximal positive skewness~\citep{RoyalChains}. This is because inertia-less fibres, once trapped inside a vortical zone, have no tendency to escape it and remain trapped until the lifetime of the vortex. Additionally, they have a higher likelihood to get trapped as their elasticity pulls them inside a vortex when one is encountered~\cite{PicardoPRL,RoyalChains}. An inertial fibre, however,  is centrifugally expelled from the strongly vortical regions so that they spend less time in the vortical regions. This means that the $\q$-pdfs show shrunk right tails with increasing St in the (main panel of) Fig.~\ref{Qpdf1}. 
This is also accompanied by a marginal shrinking of the left tails showing an undersampling of straining regions with increasing inertia. This is because fibres can follow the local flow closely when their inertia is small. As a result, they cannot exit the straining zones easily as the fluid drag forces stretch the fibre in opposing directions. However, as inertia increases, fibres are unable to follow the local flow closely and react to it only with a finite lag (of $\mathcal{O}(\tau_p))$. This results in fibres spending lesser times in the straining regions and exit them faster compared to less inertial fibres. Therefore, upto an inertia of  St $\sim \mathcal{O}(1)$, fibres increasingly undersample both straining as well as vortical regions. Clearly, they tend to stick more in the quiescent regions where $\q \approx 0$. A further increase in inertia beyond St $\approx 1$ means that the fibres lag the local flow even more and essentially decorrelate from it. Therefore,  fibres at large inertia, i.e. St $\gg 1$, populate the flow more homogeneously (countering the vortical expulsion). This is seen in the marginal widening of the right tails of $\q$-pdfs at large St $(=3.0, 8.0)$ in the main panel of Fig.~\ref{Qpdf1}. 

Elasticity of the fibres, however, counteracts the effect of inertia by trapping the fibres inside the vortices. Elastic forces tend to restore the equilibrium length of the fibres. This means the beads of the fibres are drawn closer to each-other by the spring forces. Our elastic fibres can thus be drawn (partially or entirely) inside the vortices. As a result, longer fibres have a higher tendency to partially or wholly sample a vortical region. This gives the $\q$-distributions their wider right tails at large Wi in the inset Fig.~\ref{Qpdf1}. Additionally, an increase in elasticity also results in a mild oversampling of the straining regions. This is because longer fibres in straining zones have opposing drag forces acting at their two ends and as a result as long fibres cannot escape these regions easily. This gives the mildly broader left tails to the $\q$-PDFs with increasing Wi in the inset of Fig.~\ref{Qpdf1}. 

Clearly, the combination of inertial and elastic effects makes the dynamics of a fibre complex and results in distinct preferential sampling in different regimes of St and Wi. Preference for different flow regions can be even better seen by studying the underlying chaoticity of the sampled flow regions. A measure of the degree of chaoticity is the associated Lyapunov Exponent (LEs). We discuss chaoticity and LEs in the following section.

\section*{Lyapunov Exponents and Chaoticity }
\label{Sect: Lyapunov}
\begin{figure}
	\centering
	\includegraphics[width=0.5\columnwidth]{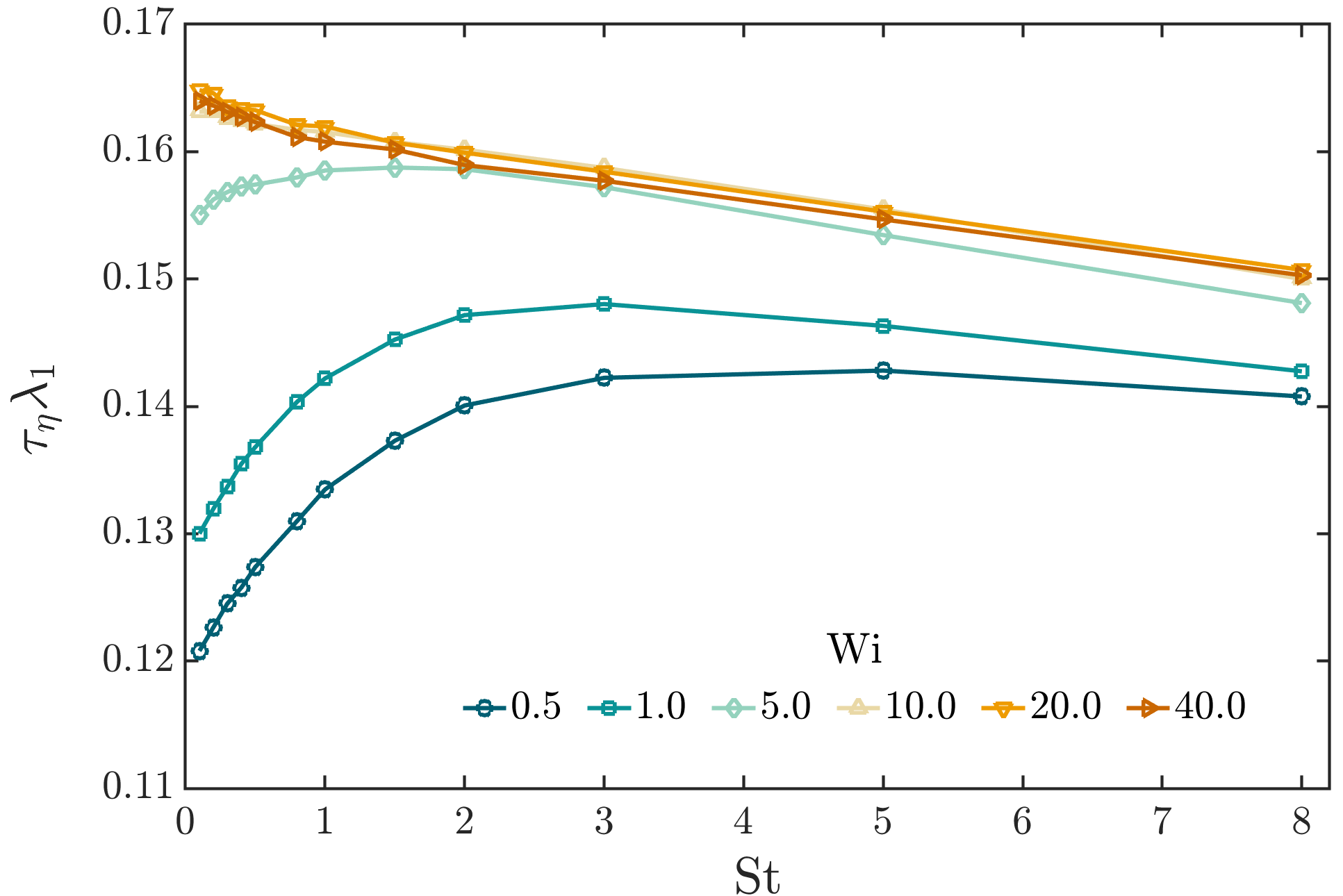}
	\caption[\textbf{Chaoticity in Trajectories of Inertial Filaments}]{Maximal Lyapunov exponents $\lambda_1$ associated with the flow regions sampled by the fibre \com es in a 3D turbulent flow. More elastic fibres sample larger values of $\lambda_1$, indicating an increased preference for regions with large stretching rates: straining zones. At large inertia, however, fibres decouple from the flow and sample it more uniformly, resulting in a dip in average sampled $\lambda_1$ values. This picture of LEs provides an alternate description to preferential sampling described previously using the $\q$-values. The non-monotonic dependence of $\lambda_1$ on St is already known for free inertial particles~\cite{Bec06}.}
	\label{LE1}
\end{figure}
 That fibres preferentially visit very different regions of a 3D turbulent flow as a function of St and Wi can also be seen in the chaoticity of the flow regions they sampled. A simple way to quantify chaoticity is the Lyapunov Exponent (LE). LEs give the typical rate of divergence (compared to an exponential) of initially nearby trajectories~\citep{Lyapunov92}. and are related to the instantaneous, local stretching rates of the velocity field (velocity gradients)~\cite{Guillaume02,Bec06,SSRLE,Meneveau15}.

  To quantify the degree of chaoticity of the sampled flow regions, we compute the LEs associated with the instantaneous fluid motion sampled by the fibre \com es. The LEs are given by the singular values of the deformation tensor $\bD$ or equivalently the square root of the eigenvalues of the Cauchy tensor $C_{ij} = D_{ik} D_{kj}$ where $D_{ij}$ are the elements of the matrix $\bD$. The evolution of $D_{ij}$ can be obtained by differentiating the evolution equation for a fluid particle $d X_i / dt = u_i({\bf X}, t)$ with respect to the coordinate~\cite{Guillaume02,Meneveau15}, where {\bf X} is the instantaneous position of the said particle:
 \begin{align}
 	\frac{d \Dij}{dt} = \left( \delu{i}{k} \right)D_{kj} \ , \quad \Dij (t=0) = \delta_{ij},			\label{DefTen}
 \end{align}
 The above equation describes how a local velocity gradient deforms a (initial) unit sphere by stretching/compressing it in different directions. The semi-axes of the resulting ellipse are given by the singular values $\sigma_i(\bx,t)$ of the deformation tensor $\bD(t)$. The exponential growth rate of these singular values yields the Finite Time Lyapunov Exponents (FTLEs) $\gamma_i$, $i = 1,2,3$:
 \begin{align}
 	\gamma_i (\bx,t) = \frac{1}{t-t_0} \ln \sigma_i (\bx,t)
 \end{align}
 LEs are obtained via the ensemble average of FTLEs in the long time limit~\cite{Meneveau15}:
 \begin{align}
 	\lambda_i = \langle \gamma_i \rangle = \lim_{t-t_0 \to \infty} \frac{1}{t-t_0} \langle \ln \sigma_i \rangle
 \end{align}
 The maximal LE $\lambda_1$ is an indicator of the chaoticity of the local flow regions. The $\lambda_1$'s (non-dimensionalized by $\tau_\eta$) thus computed are plotted in Fig.~\ref{LE1} as a function of inertia and elasticity. Our bead-spring fibres, for very small elasticity, are very stiff and prefer the vortical regions much more than the straining zones of the flow. This is because small, stiff springs are more likely to be completely drawn inside vortices and experience correlated drag forces along their lengths in the straining zones. This makes their exit faster allows for only a minimal sampling of large stretching rates. As a result, fibres have smaller associated maximal LEs $\lambda_1$ for small Wi, see Fig.~\ref{LE1}. Less elastic fibres, thus, prefer less chaotic regions of the flow. Now, as inertia is increased, fibres begin to sample the straining zones more as a result of vortical ejection. As these regions are marked by typically large stretching rates, we have larger $\lambda_1$ at moderate St. At even larger inertia, they begin to decorrelate from the flow and, as a result, populate the flow more uniformly. This decreases $\lambda_1$ on the average as they can now approach the vortical regions more closely. Hence, the dip in the $\lambda_1$-St curve at large St. 
 
 With increasing Wi, fibres with small inertia oversample both straining as well the vortical regions as discussed in the previous section (see inset of  Fig.~\ref{Qpdf1}). The large stretching rates in the straining regions (in addition to contributions due to vortex stretching) means that the fibre \com es sample larger values of $\lambda_1$. This is exactly what we see in Fig.~\ref{LE1}, especially for small St. As these large Wi fibres become more inertial, they decouple from the flow and sample it more uniformly. 
 This is manifested in a monotonic decrease in $\lambda_1$ with St when fibre elasticity is large. Thus, it is clear that fibres with large elasticity prefer more chaotic regions of the flow. 
 
 \section*{Conclusions and Summary}
 
 In conclusion, we have shown that the stretching of fibres is due only to elasticity and their inertia playing a minimal role as they are advected by a turbulent carrier flow. A highly elastic fibre is much more likely to be stretched out and as a result prefers a ``straighter" configuration rather than a coiled one. These inertial, elastic fibres then exhibit non-trivial preferential sampling of a 3D turbulent flow in a manner qualitatively similar to 2D turbulence~\cite{SinghPRE}. Inertia leads fibres away from vortical regions while their elasticity pulls them inside the vortices. Upto a moderate inertia (St $\sim \mathcal{O}(1)$), fibres increasingly prefer the straining regions of the flow, while at much larger inertia (St $\gg 1$) they decorrelate from the flow and preference for straining regions begins to diminish again. However, owing to a large elasticity, fibres get trapped in vortical regions (at small St), as well as are unable able to exit the straining regions quickly. A more elastic and extensible fibre is, thus, more likely to spend longer times in both vortical and the straining regions of the flow. 
 
 This picture of preferential sampling of a 3D turbulent flow by elastic, inertial fibres is also confirmed by alternately studying the chaoticity of the sampled flow regions via Lyapunov Exponents. Less elastic fibres prefer less chaotic (vortical) regions of the flow while more chaotic (straining) regions are preferred at large Wi. LEs also confirm that preferential sampling has a non-monotonic dependence on St for small elasticity but which is lost when Wi becomes very large. This approach to preferential sampling via Lyapunov Exponents not only confirms our understanding but also captures the underlying nature of different flow regions sampled by the fibres. It would, however, be even more interesting to see how chaotic the fibre trajectories themselves are and what that has to say about fibre dynamics in turbulent flows.

\section*{Acknowledgment}
	 RKS thanks Samriddhi Sankar Ray for useful discussions and suggestions on the manuscript. 
	 The simulations were performed on the ICTS clusters Contra and Tetris as well as the work stations from the project ECR/2015/000361: Goopy and Bagha. Since a part of this work was done during his PhD, RKS acknowledges support of the DAE, Govt. of India, under project no. 12-R\&D-TFR-5.10-1100 and SSR SERB-DST (India) projects MTR/2019/001553, STR/2021/000023 and CRG/2021/002766 for financial support.
	
	\bibliography{ref_turb_fibres}
\end{document}